\documentclass[runningheads]{llncs}

\usepackage[T1]{fontenc}
\usepackage{amsmath,amssymb,amsfonts}
\usepackage{mathtools}
\usepackage{graphicx}
\usepackage{textcomp}
\usepackage{xcolor}
\usepackage{array,tabularx}
\usepackage{pifont}
\usepackage{physics}

\newcolumntype{L}[1]{>{\raggedright\arraybackslash}p{#1}}
\usepackage{algorithm}
\usepackage{algpseudocode}
\usepackage{listings}
\usepackage{comment}
\usepackage{tikz}
\usetikzlibrary{positioning, fit, arrows.meta}

\usepackage[hidelinks]{hyperref}

\providecommand{\Letter}{\ding{41}}
\providecommand{\discintname}{Disclosure of Interests.}
\makeatletter
\@ifundefined{credits}{%
  \newenvironment{credits}{\begingroup\let\subsubsection\paragraph}{\endgroup}%
}{}
\makeatother

\begin{document}

\title{A Simulation Framework for Noise and Error-Correction Analysis in Device-Independent Quantum Key Distribution}

\titlerunning{A Simulation Framework for Noise and Error-Correction Analysis in DIQKD}

\author{%
Nguyen Duong Hoang Duy\inst{1,2} \and
Nguyen Trinh Dong\inst{3} \and
Vu Tuan Hai\inst{1,2} \and
Le Vu Trung Duong\inst{4} \and
Nguyen Van Tinh\inst{5}\textsuperscript{(\Letter)}%
}

\authorrunning{N. D. H. Duy et al.}

\institute{%
University of Information Technology, Ho Chi Minh City, Vietnam \and
Vietnam National University, Ho Chi Minh City, Vietnam \and
Faculty of Computer Science and Engineering, Thuyloi University, Hanoi, Vietnam \and
Nara Institute of Science and Technology, Nara, Japan \and
Le Quy Don Technical University, Hanoi, Vietnam\\
\email{take@lqdtu.edu.vn}%
}

\maketitle

\begin{abstract}
Device-Independent Quantum Key Distribution (DIQKD) certifies security from the violation of a Bell inequality alone, but its practical margin against noise is narrow and is usually quoted only through the Bell violation itself. We build a
simulator of the event-ready DIQKD experiment and use it to trace readout noise through the Bell test into classical
post-processing. For a symmetric outcome-flip channel of strength $e$, we reproduce the simulated violation window $e<5.86\%$, to within $0.11$ percentage points. Combining these with a multi-pass Cascade reconciliation and the Devetak-Winter bound gives $r \geq 0.162$ bits per round, and shows that a positive key rate requires $e < 0.89\%$ - roughly one-sixth of the window in which a Bell violation is still observed. The gap quantifies how far certified
nonlocality overstates the usable noise budget of a DIQKD link.

\end{abstract}

\section{Introduction}

Quantum communication distributes quantum information over insecure channels \cite{Primaatmaja2023securityofdevice}. Its primary application, Quantum Key Distribution (QKD), lets two users (Alice and Bob) establish a shared secret key over an insecure quantum channel and an authenticated classical one.
Unlike classical cryptography, which relies on high computational complexity problems, QKD provides information-theoretic security guaranteed by physics. The field began
with the Prepare-and-Measure (P\&M) scheme, most notably BB84
\cite{bennett1984quantum}, followed by entanglement-based schemes such as the Ekert protocol \cite{ekert1991quantum} and BBM92 \cite{bennett1992quantum}. More recent work includes Measurement-Device-Independent (MDI) QKD \cite{goh2016measurement}, in which an untrusted third party performs the measurement, closing detector-related side channels. DIQKD goes further: security is certified without any model of the internal workings of the devices, which are treated as black boxes mapping inputs to outputs. Certification rests on Bell nonlocality \cite{brunner2014bell}, a higher violation implying weaker correlation with an adversary through the
monogamy of nonlocality. The idea traces to Ekert's use of Bell nonlocality \cite{ekert1991quantum}, with self-testing formalized later \cite{mayers1998quantum,acin2007device}.

Between the Bell test and a usable key lies the classical post-processing layer. Information reconciliation turns the two correlated but potentially different raw keys into an identical string, and privacy amplification compresses that string to a secret
one. Both steps consume the entropy certified by the Bell test, so the violation by itself does not determine whether a link is operable. Because DIQKD hardware is scarce, this end-to-end accounting is most readily studied in simulation,
where the device can still be treated as a black box and results certified from measurement data alone.

This paper makes three contributions. (1) We build a DIQKD simulator implementing the experimental settings and event-ready entanglement swapping scheme of \cite{zhang2022device} and reproduce its reported operating point. (2) We derive
closed-form expressions for the CHSH value and the QBER under a symmetric outcome-flip channel, and confirm them against the simulated noise sweep; they explain the observed violation window and its symmetry about $e=1/2$. (3) We integrate an explicit multi-pass Cascade reconciliation and evaluate the
Devetak-Winter bound, obtaining the secret key rate at the operating point and the noise threshold beyond which it vanishes - a threshold roughly six times tighter than the Bell-violation threshold.

\begin{figure}[t]
    \centering
    \includegraphics[width=\textwidth]{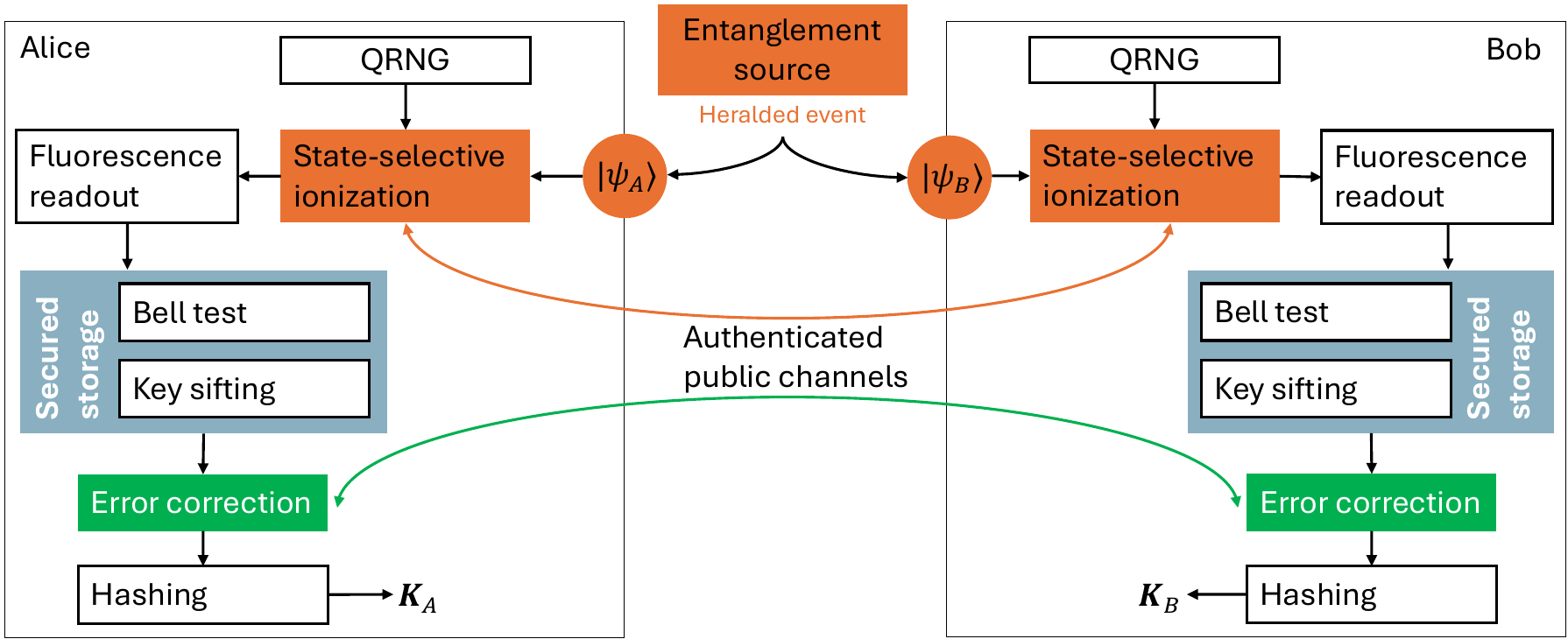}
    \caption{Simulated DIQKD protocol system. An entanglement source generates
    correlated photon pairs sent to Alice and Bob. Each party uses a Quantum
    Random Number Generator (QRNG) to choose measurement settings and measures
    its photon, with outcomes obtained through state-selective ionization and
    fluorescence readout, then stored locally. Results are exchanged over an
    authenticated channel, where a Bell test verifies entanglement and detects
    eavesdropping. Key sifting then retains compatible outcomes, error
    correction reconciles discrepancies, and privacy amplification (universal
    hashing) removes leaked information, producing identical secure keys}
    \label{fig:scheme}
\end{figure}

\section{Related Work}

\begin{table}[t]
\centering
\caption{Thresholds for variants of the CHSH-based DIQKD protocol. The upper
block reports values established analytically in the cited work. }
\label{tab:related_work}
\renewcommand{\arraystretch}{1.2}
\resizebox{\textwidth}{!}{%
\begin{tabular}{llll}
\hline
\textbf{Protocol} & \textbf{QBER} & \textbf{Critical QBER} & \textbf{Critical Detection Efficiency} \\ \hline
The standard protocol & -- & 7.1\% \cite{pironio2009device} & 86.5\% \cite{woodhead2021device,Sekatski2021deviceindependent} \\
Generalized CHSH (Gen. CHSH) & -- & 7.4\% \cite{woodhead2021device,Sekatski2021deviceindependent} & 86.5\% \cite{woodhead2021device,Sekatski2021deviceindependent} \\
Noisy pre-processing (NPP) & -- & 8.1\% \cite{woodhead2021device,Sekatski2021deviceindependent} & 82.6\% \cite{woodhead2021device,Sekatski2021deviceindependent}\\
Random key-basis (RKB) & -- & 8.4\% \cite{Masini2022simplepractical} & 92.5\% \cite{Schwonnek2021} \\
Random post-selection & -- & -- & 68.5\% \cite{Primaatmaja2023securityofdevice} \\
Advantage distillation & -- & 9.1\% \cite{Tan2021} & 89.1\% \cite{Tan2021} \\
NPP + Gen. CHSH (+ bias) & -- & 8.3\% \cite{woodhead2021device,Sekatski2021deviceindependent} & 80.3\% \cite{Masini2022simplepractical} \\
RKB + NPP & -- & 9.3\% \cite{Tan2022improveddiqkd} & -- \\
Heralded entanglement + RKB (Baseline) & 7.8\% \cite{zhang2022device} & 8.2\% \cite{zhang2022device} & 100\% \cite{zhang2022device} \\
\hline
Our simulation & \textbf{7.7\%} & -- & -- \\
\hline
\end{tabular}
}
\end{table}

\subsection{Hardware-Based Methods}

Practical DIQKD is limited by detection loopholes and channel losses
\cite{pearle1970hidden}, as represented in Figure~\ref{fig:scheme}. Hardware
advances addressing these include superconducting nanowire single-photon
detectors (SNSPDs) exceeding $87\%$ efficiency \cite{liu2022toward}, and qubit
amplifiers based on teleportation or entanglement swapping that herald photon
detection without disturbing the state \cite{gisin2010proposal}. Heralded
entanglement between matter qubits such as trapped ions or neutral atoms
achieves high-fidelity measurement and strong Bell violation at low QBER
\cite{nadlinger2022experimental}, while ultra-low-loss fibers (approaching
$0.2$~dB/km) extend the range over which positive key rates are attainable.

\subsection{Software-Based Methods}

Protocol-level improvements act on post-processing and measurement strategy.
Noisy pre-processing lowers the detection-efficiency requirement by randomly
flipping key bits, reducing an eavesdropper's information
\cite{woodhead2021device}; advantage distillation uses two-way communication to
raise the tolerable QBER to about $9.1\%$ \cite{tan2020advantage}; randomized key
bases distribute measurement uncertainty \cite{pironio2009device}; and random
post-selection strengthens correlations by discarding selected events
\cite{liu2022toward}. Asymmetric configurations and full-behaviour analyses
exploit the complete correlation statistics to improve noise resilience
\cite{woodhead2021device}.

\subsection{Algorithmic Analysis}

Numerical methods and security proofs have tightened DIQKD key-rate bounds,
particularly in the finite-key regime. The Entropy Accumulation Theorem (EAT)
relates per-round von Neumann entropy to the total smooth min-entropy under
general attacks \cite{dupuis2019entropy}, while semi-definite programming such
as the NPA hierarchy yields systematic entropy bounds for uncharacterized
devices \cite{navascues2007bounding}. Improved entropy inequalities sharpen
conditional-entropy estimates beyond methods based on guessing probability
\cite{tan2020advantage}, and Quantum Probability Estimation (QPE) reduces the
number of rounds needed for certification \cite{knill2018quantum}.

\section{Device-Independent Quantum Key Distribution Protocol}

\begin{figure}[t]
    \centering
    \includegraphics[width=0.72\textwidth]{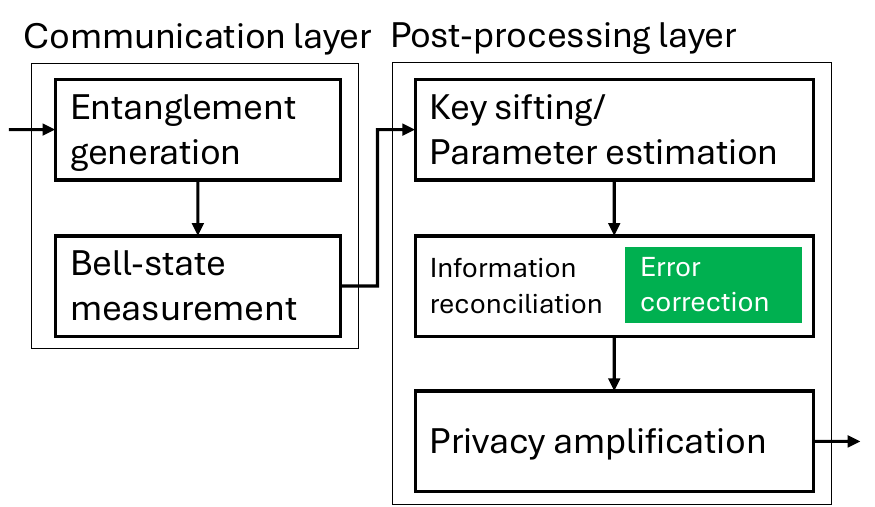}
    \caption{Standard DIQKD. In this work, Cascade error correction is
    incorporated into the information reconciliation stage of the classical
    post-processing layer}
    \label{fig:standard_diqkd}
\end{figure}

DIQKD certifies security from uncharacterized ``black-box'' devices, using Bell
nonlocality to remove device modelling and mitigate side-channel attacks. It
assumes the validity of quantum mechanics
\cite{acin2007device,pironio2009device}, secure laboratories with trusted
classical components \cite{vazirani2014fully}, an authenticated classical channel
\cite{renner2008security}, and private randomness \cite{colbeck2011free}.
Loophole-free Bell tests - closing the detection, locality and
measurement-dependence loopholes - are required to prevent adversarial
simulation of nonlocal correlations
\cite{hensen2015loophole,giustina2015significant,shalm2015strong}. The protocol
splits into a quantum communication layer and a classical post-processing layer
(Figure~\ref{fig:standard_diqkd}).

\begin{table}[t]
\centering
\caption{Twenty rounds of the standard protocol of Pironio et al.
\cite{pironio2009device}, whose convention ($x \in \{0,1\}$, $y \in \{0,1,2\}$,
key rounds at $(x,y)=(0,2)$) is used here for illustration; the settings of our
simulator are those of \cite{zhang2022device} and are given in
Section~\ref{sec:method}. Each round draws inputs $x,y$ and yields outcomes
$a,b \in \{-1,+1\}$. Only test rounds enter the CHSH estimate and only key rounds
produce the raw key bits $A, B$, so blank cells are rounds of the other type.
Outcomes map to bits by $+1 \mapsto 1$, $-1 \mapsto 0$. Bold entries mark the key
rounds where $A \neq B$}
\label{tab:example_qkd}
\setlength{\tabcolsep}{2pt}
\renewcommand{\arraystretch}{1.2}
\resizebox{\textwidth}{!}{%
\begin{tabular}{lcccccccccccccccccccc}
\hline
\textbf{\#Rounds}
& \textbf{1} & \textbf{2} & \textbf{3} & \textbf{4} & \textbf{5} & \textbf{6} & \textbf{7} & \textbf{8} & \textbf{9} & \textbf{10}
& \textbf{11} & \textbf{12} & \textbf{13} & \textbf{14} & \textbf{15} & \textbf{16} & \textbf{17} & \textbf{18} & \textbf{19} & \textbf{20} \\
\hline
type
& test & test & key & key & key
& key & test & test & test & test
& key & test & key & test & test
& key & key & key & test & test \\
$x$
& 1 & 1 & 0 & 0 & 0
& 0 & 1 & 1 & 1 & 1
& 0 & 1 & 0 & 1 & 1
& 0 & 0 & 0 & 1 & 1 \\
$a$
& +1 & +1 & +1 & +1 & -1
& +1 & -1 & -1 & -1 & +1
& +1 & -1 & +1 & -1 & +1
& -1 & -1 & +1 & -1 & +1 \\
$y$
& 0 & 0 & 2 & 2 & 2
& 2 & 1 & 1 & 0 & 1
& 2 & 0 & 2 & 1 & 1
& 2 & 2 & 2 & 1 & 1 \\
$b$
& -1 & -1 & -1 & +1 & -1
& -1 & -1 & +1 & +1 & -1
& -1 & -1 & +1 & -1 & -1
& -1 & -1 & -1 & +1 & +1 \\
A
&  &  & \textbf{1} & 1 & 0
& \textbf{1} &  &  &  &
& \textbf{1} &  & 1 &  &
& 0 & 0 & 1 &  &  \\
B
&  &  & \textbf{0} & 1 & 0
& \textbf{0} &  &  &  &
& \textbf{0} &  & 1 &  &
& 0 & 0 & 1 &  &  \\
\hline
\end{tabular}}
\end{table}

\paragraph{Quantum communication layer.}
In each round, Alice and Bob share a bipartite state $\rho_{AB}$ from a source
that may be untrusted. Alice draws a measurement input $x \in X$ and Bob
$y \in Y$, yielding binary outcomes $a,b \in \{-1,+1\}$. Key rounds use
$x \in Rem_A$, $y \in Rem_B$ and produce the raw key; test rounds use
$x \in Sub_A$, $y \in Sub_B$ and estimate the nonlocal correlations.

\paragraph{Classical post-processing layer.}
Alice and Bob announce their inputs to separate key from test rounds, then
reveal a subset of the test data to compute the CHSH value
\begin{align}\label{eq:chsh}
    S \coloneqq \langle A_{x_0} B_{y_0} \rangle
+ \langle A_{x_0} B_{y_1} \rangle
+ \langle A_{x_1} B_{y_0} \rangle
- \langle A_{x_1} B_{y_1} \rangle ,
\end{align}
with $x_0,x_1 \in Sub_A$ and $y_0,y_1 \in Sub_B$. The correlators
$\langle A_x B_y \rangle$ are averages of the outcome products over rounds with
the same input pair \cite{Primaatmaja2023securityofdevice}. Local
hidden-variable models satisfy $S \le 2$, whereas quantum mechanics permits
$S \le 2\sqrt{2}$, Tsirelson's bound \cite{clauser1969proposed}. The protocol
aborts if $S \leq 2$.

Otherwise the parties sift, retaining generation rounds with matching settings,
and apply the mapping $\mathcal{M}:\{-1,+1\}\rightarrow\{0,1\}$ to obtain sifted
keys $K^*_A, K^*_B$ with $Q = \Pr[K^*_A \neq K^*_B]$. Table~\ref{tab:example_qkd}
illustrates twenty such rounds. Information reconciliation then makes the two
strings identical and verifies the result by comparing short two-universal hash
digests, guaranteeing correctness with probability $\epsilon_{cor}$
\cite{portmann2014cryptographicsecurityquantumkey}. Privacy amplification
finally hashes the reconciled string to a shorter key; by the Quantum Leftover
Hash Lemma, a length below the smooth min-entropy guarantees secrecy
$\epsilon_{sec}$ and near-uniform randomness \cite{tomamichel2011leftover}.

\section{Proposed Method}\label{sec:method}

\subsection{DIQKD Simulation}

\begin{algorithm}[t]
\caption{DIQKD simulation.
The source code is available at \cite{githubGitHubDuyndh98DIQKD}.}
\label{algo:1}
\begin{algorithmic}
\Require \#Rounds $N \in \mathbb{N}$, key round settings $Rem_A, Rem_B$, test round settings $Sub_A, Sub_B$
\Ensure CHSH $S$, QBER $Q$, secret key $K$
\State Initialize $x_i, y_i, a_i, b_i \gets 0, \forall i \in \{1,\dots,N\}$
\For{$i \gets 1$ to $N$}
    \State $\lambda_i \gets \mathcal{E}()$ \Comment{entanglement generation}
    \State $(x_i, y_i) \gets \mathcal{R}()$ \Comment{random inputs}
    \State $(a_i^{*}, b_i^{*}) \gets \mathcal{S}(\lambda_i, x_i, y_i)$ \Comment{state-selective ionization}
    \State $(\hat{a_i}, \hat{b_i}) \gets \mathcal{F}(a_i^{*}, b_i^{*})$ \Comment{fluorescence readout}
    \State $(a_i, b_i) \gets (\mathcal{M_A}(\hat{a_i}), \mathcal{M_B}(\hat{b_i}))$ \Comment{mapping output}
\EndFor
\State $I_\mathcal{T} \gets \{ i \mid x_i \in Sub_A, y_i \in Sub_B\}$ \Comment{test rounds}
\State $I_\mathcal{K} \gets \{ i \mid x_i \in Rem_A, y_i \in Rem_B \}$ \Comment{key rounds}
\State $I_\mathcal{K^*} \gets \{i \mid i \in I_\mathcal{K}, x_i = y_i\}$ \Comment{sifted key rounds}
\State $S \gets CHSH\big(\{(x_i,y_i,a_i,b_i)\}_{i \in I_\mathcal{T}}\big)$ \Comment{CHSH on test rounds}
\State $(K^*_A, K^*_B) \gets \{(a_i,b_i)_{i \in I_\mathcal{K^*}}\}$ \Comment{sifted key}
\State $Q \gets ({1}/{|I_\mathcal{K^*}|}) \sum_{i \in I_\mathcal{K^*}} [a_i \oplus b_i]$ \Comment{QBER on sifted key}
\State $(K_A', K_B') \gets ErrorCorrection(K^*_A, K^*_B)$ \Comment{Cascade (Algorithm~\ref{alg:cascade})}
\State $h \gets \mathcal{H}$ \Comment{random 2-universal hash function}
\State $(K_A, K_B) \gets (h(K_A'),\, h(K_B'))$ \Comment{hashing}
\State \textbf{Result:} $S$, $Q$, $K = K_A = K_B$
\end{algorithmic}
\end{algorithm}

We implement the simulator from scratch, following the components and physical
interactions of the DIQKD system of \cite{zhang2022device}
(Algorithm~\ref{algo:1}). Laboratory setups are treated as uncharacterized
black boxes, so security is certified from input-output data alone, and the
program models event-ready entanglement swapping between two remote rubidium
atoms connected by optical fiber.

Device parameters and measurement settings follow the original experiment
\cite{zhang2022device}. Alice's four inputs $X \in \{0,1,2,3\}$ select the
orientation angles $-22.5^\circ$, $+22.5^\circ$, $-45^\circ$ and $0^\circ$,
and Bob's two inputs $Y \in \{0,1\}$ select $+22.5^\circ$ and $-22.5^\circ$. Key
rounds use $Rem_A = Rem_B = \{0,1\}$ and test rounds $Sub_A = \{2,3\}$,
$Sub_B = \{0,1\}$. Bob's outcome mapping is
$\mathcal{M}_B(x) = (x + 1) \gg 1$, and Alice applies the inverted map
$\mathcal{M}_A(x) = 1 - \mathcal{M}_B(x)$ to compensate for the anti-correlations
of $\ket{\Psi^+}$. With the reported entanglement fidelity of $\geq 0.892(23)$,
the simulator produces $S = 2.579$ and an average QBER of $0.077$, against
$7.8\%$ measured in \cite{zhang2022device}; $8.2\%$ is taken as the reference
operating point.

\subsection{Noise Model}\label{sec:noise}

Readout imperfections such as magnetic-field fluctuations and ionization
inefficiencies are modeled as a symmetric flip applied independently to each
recorded outcome with probability $e$. Writing the flip as a sign
$s \in \{-1,+1\}$ with $\mathbb{E}[s] = 1-2e$, independence of the two readouts
gives $\langle a'b'\rangle = \mathbb{E}[s_A]\,\mathbb{E}[s_B]\,\langle ab\rangle$,
so every correlator is scaled by the same factor and
\begin{align}\label{eq:model}
&S(e) = (1-2e)^2\,S(0), \\
&Q(e) = Q(0) + \big(1-2Q(0)\big)\,2e(1-e).
\end{align}
Both depend on $e$ only through $(1-2e)^2$ and $e(1-e)$, hence are symmetric
about $e = 1/2$: the values at $e$ and $1-e$ differ by a deterministic
relabelling of both records, so $e \in [0,1/2]$ is the physically distinct
range. Equation~(\ref{eq:model}) predicts a Bell violation for
$e < \tfrac{1}{2}\big(1-\sqrt{2/S(0)}\big) = 5.97\%$ and, through $Q(e)$.

\subsection{Cascade Error Correction}

\begin{algorithm}[t]
\caption{The Cascade error correction}\label{alg:cascade}
\begin{algorithmic}
\Require Strings $X, Y \in \{0, 1\}^N$, block sizes $\{k_i\}_{i=1}^M$, random functions $\{f_i\}_{i=1}^M$
\Ensure $X = Y$
\For{$i \gets 1$ to $M$}
    \State Define blocks $K_j^i = \{l \mid f_i(l) = j\}$ for $j \in \{1, \dots, N/k_i\}$
    \For{each block $K_j^i$}
        \State Alice sends $P(A, i, j) \gets \bigoplus_{l \in K_j^i} X_l$
        \State Bob computes $P(B, i, j) \gets \bigoplus_{l \in K_j^i} Y_l$
        \If{$P(A, i, j) \neq P(B, i, j)$}
            \State $l \gets \text{BINARY}(K_j^i)$
            \If{$i > 1$}
                \State $\mathcal{L} \gets \{ (m, n) \mid m < i, l \in K_n^m \}$ \While{$\mathcal{L} \neq \emptyset$}
                    \State $(m^*, n^*) \gets \arg\min_{(m, n) \in \mathcal{L}} |K_n^m|$ \If{$\bigoplus_{k \in K_{n^*}^{m^*}} X_k \neq \bigoplus_{k \in K_{n^*}^{m^*}} Y_k$}
                        \State $l' \gets \text{BINARY}(K_{n^*}^{m^*})$
                        \State $\mathcal{L} \gets \mathcal{L} \cup \{ (m, n) \mid m < i, l' \in K_n^m \}$
                    \EndIf
                    \State $\mathcal{L} \gets \mathcal{L} \setminus \{(m^*, n^*)\}$
                \EndWhile
            \EndIf
        \EndIf
    \EndFor
    \State \textbf{Result:} $\forall m \le i, \forall n: \bigoplus_{l \in K_n^m} X_l = \bigoplus_{l \in K_n^m} Y_l$
\EndFor
\end{algorithmic}
\end{algorithm}

Cascade is a two-way interactive protocol that reconciles bit strings by
recursive BINARY search \cite{tupkary2023using}. The data are partitioned into
blocks over several passes; a parity mismatch triggers localisation and
correction of one error, and corrections in later passes expose inconsistencies
in earlier blocks, producing the ``cascade effect'' of iterative backtracking
until all parities agree. We apply the full multi-pass procedure to the sifted
keys rather than assuming a fixed efficiency, and obtain the final key by
accounting for the disclosed information and applying privacy amplification.

\subsection{Secret Key Rate}

We evaluate asymptotic security against collective attacks through the
Devetak-Winter bound
\begin{equation}\label{eq:keyrate}
r \ge 1 - h\!\left( \frac{1 + \sqrt{(S/2)^2 - 1}}{2} \right) - f_{\text{EC}} \cdot h(Q),
\end{equation}
where $h(p) = -p \log_2 p - (1-p)\log_2(1-p)$ and $f_{\text{EC}} \geq 1$ is the
reconciliation efficiency, the ratio of the information disclosed during
reconciliation to the Shannon limit $h(Q)$. At the reproduced operating point
($S = 2.579$, $Q = 0.077$) the two entropy terms are $0.446$ and $0.392$, giving
\begin{align}\label{eq:rnum}
&r \geq 0.162 \quad (f_{\text{EC}} = 1), \nonumber\\
&r \geq 0.104 \quad (f_{\text{EC}} = 1.15),
\end{align}
bits per round, the second value using the efficiency assumed in
\cite{zhang2022device}. The rate stays positive for any
$f_{\text{EC}} < 1.415$, which covers the range reported for Cascade
\cite{tupkary2023using}, so the conclusions below do not hinge on the precise
reconciliation efficiency. Composing (\ref{eq:keyrate}) with the noise model
(\ref{eq:model}) yields $r(e)$ and hence the noise threshold for a positive key
rate, evaluated in Section~\ref{sec:results}.

\section{Experiment Results}\label{sec:results}

\begin{figure}[http]
    \centering
    \includegraphics[width=\textwidth]{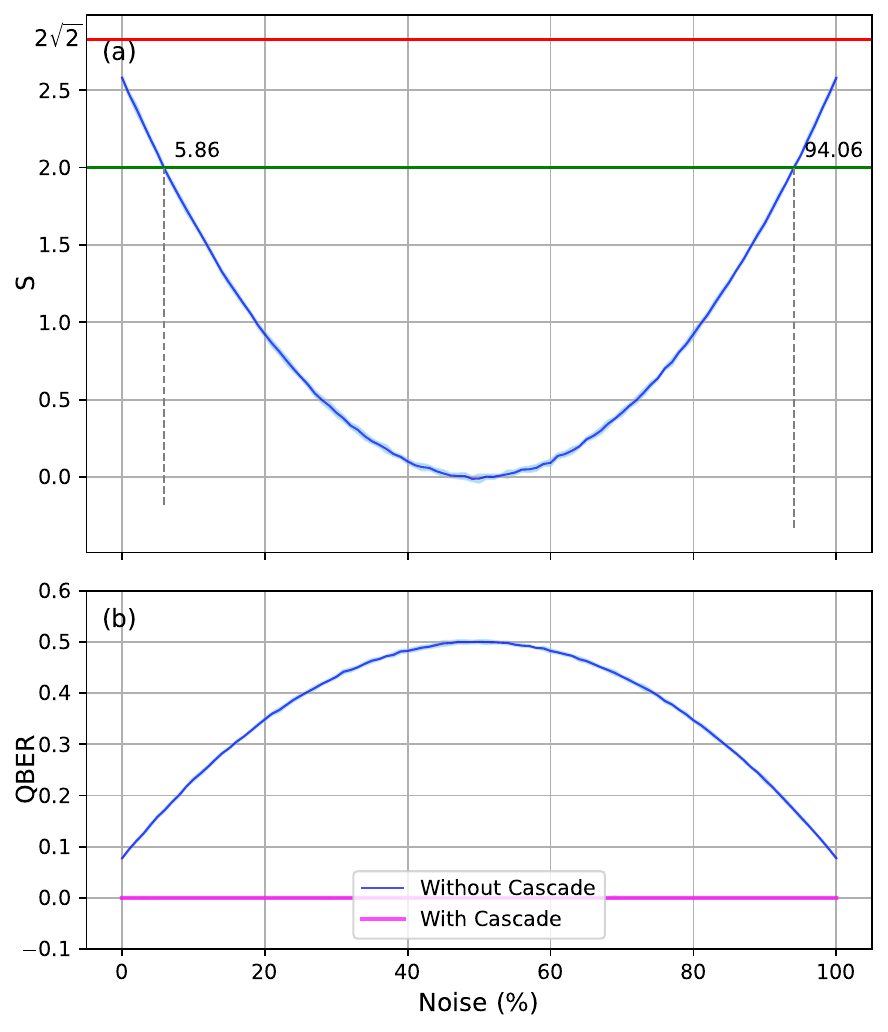}
    \caption{(a) $S$ versus outcome-flip probability $e$, from $10$ runs of
    $100{,}000$ rounds. Horizontal lines mark the classical bound $S=2$ and
    Tsirelson's bound $S=2\sqrt{2}$; violation is observed for $e<5.86\%$ and,
    by the symmetry of (\ref{eq:model}), above $94.06\%$. (b) QBER without Cascade and with Cascade.}
    \label{fig:noise_qber_cascade}
\end{figure}

Figure~\ref{fig:noise_qber_cascade}(a) shows the CHSH value across the noise
sweep, averaged over $10$ runs of $100{,}000$ rounds at each level. $S$ falls
from $2.579$ as $e$ grows, reaching zero near $e = 1/2$ where the flip
randomises each outcome and destroys the correlations, then recovers
symmetrically. The measured violation window, $e < 5.86\%$ and $e > 94.06\%$,
agrees with the thresholds $5.97\%$ and $94.03\%$ predicted by
(\ref{eq:model}) to within $0.11$ percentage points, confirming the model. The
symmetry is exact rather than incidental: the channel enters only through
$(1-2e)^2$, so $e$ and $1-e$ produce identical statistics up to a deterministic
relabelling of both records, and the two branches describe the same physics. Figure~\ref{fig:noise_qber_cascade}(b) shows the QBER. The raw curve follows
$Q(e)$ of (\ref{eq:model}), rising from $0.077$ to $1/2$ at $e = 1/2$ and
returning symmetrically, so the error profile is non-monotone even though the
underlying flip probability is not. After reconciliation, the curve is flat at
zero across the whole sweep: Cascade repeats parity exchange and BINARY search
until every block parity agrees, terminating with $K_A = K_B$, and we use this
as a correctness check on the implementation. The security-relevant cost of
reconciliation is not this residual rate but the parity bits disclosed in
reaching agreement, which enter (\ref{eq:keyrate}) through $f_{\text{EC}}$.

\begin{figure}[http]
    \centering
    \includegraphics[width=0.7\textwidth]{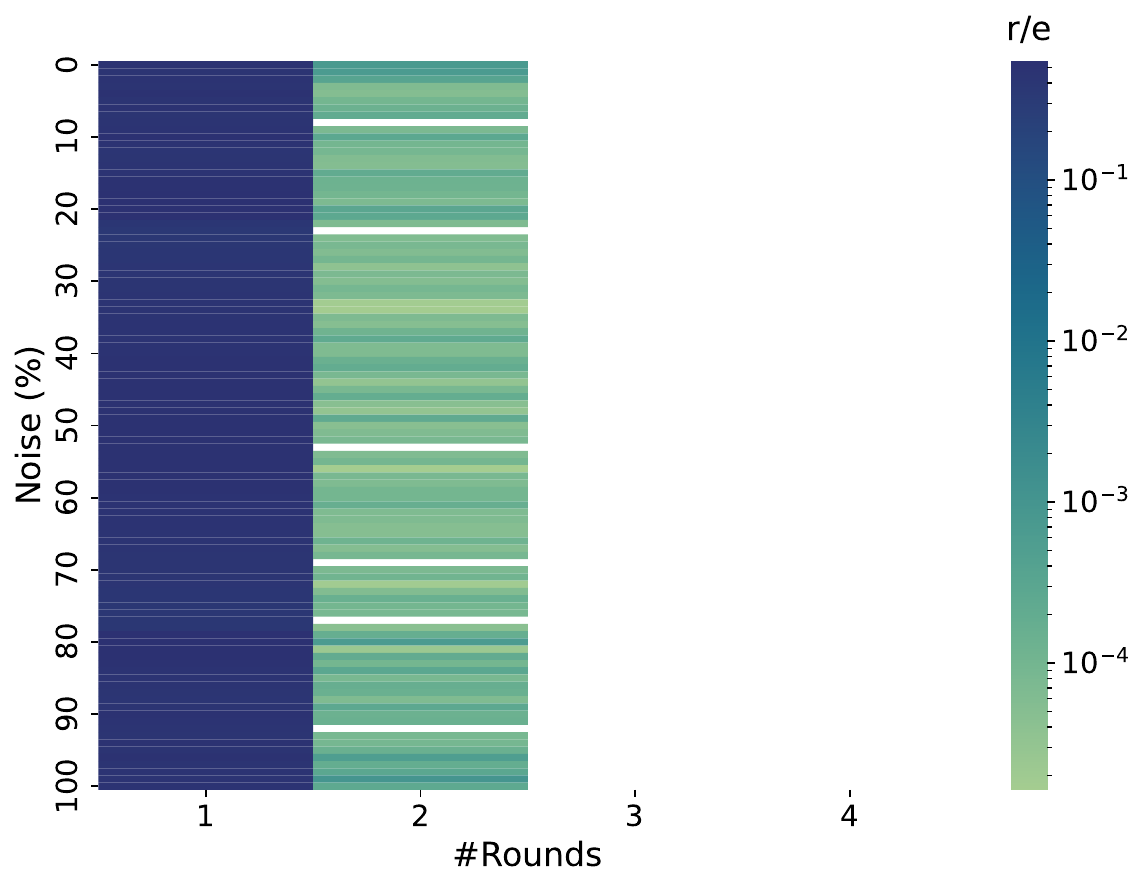}
    \caption{Remaining error ratio $r/e$ after each Cascade pass, on a
    logarithmic scale. Reconciliation converges within two passes across the
    entire noise range, so passes $3$ and $4$ contain no residual errors}
    \label{fig:cascade_heatmap}
\end{figure}

Figure~\ref{fig:cascade_heatmap} resolves this convergence pass by pass, with
initial error counts drawn from the noise sweep and Cascade run for up to four
passes. The remaining error ratio drops by about three orders of magnitude in
the first pass and reaches zero in the second, consistent with the BINARY
sub-protocol isolating one error per mismatched block while backtracking clears
the parities exposed earlier. Convergence takes the same two passes at every
noise level tested, so the number of passes is insensitive to $e$ and the
noise dependence of reconciliation resides entirely in the volume of disclosed
parities.

\begin{table}[t]
\centering
\caption{Noise budget at the reproduced operating point.}
\label{tab:budget}
\renewcommand{\arraystretch}{1.2}
\begin{tabular}{lc}
\hline
\textbf{Criterion} & \textbf{Maximum outcome-flip probability $e$} \\ \hline
Bell violation, $S > 2$ & $5.86\%$ (measured), $5.97\%$ (model) \\
Positive key rate, $f_{\text{EC}} = 1$ & $0.89\%$ \\
Positive key rate, $f_{\text{EC}} = 1.15$ & $0.53\%$ \\
\hline
\end{tabular}
\end{table}

Composing (\ref{eq:model}) with (\ref{eq:keyrate}) converts the sweep into a
noise budget, summarized in Table~\ref{tab:budget}. A positive key rate requires
$e < 0.89\%$ under ideal reconciliation and $e < 0.53\%$ at
$f_{\text{EC}} = 1.15$, against the $5.86\%$ at which the Bell violation is
still observed. Two effects compound over this interval: the first entropy term
in (\ref{eq:keyrate}) tends to $1$ as $S$ approaches the classical bound, while
$Q(e)$ grows and inflates the reconciliation cost. The usable noise budget of a
DIQKD link is therefore roughly one sixth of what the Bell violation alone
suggests - a margin that matters when the violation is the quantity most
readily measured in the laboratory.

\section{Conclusion}

We built a DIQKD simulator and traced readout noise through the Bell test into classical post-processing. A symmetric outcome-flip channel of strength $e$ scales every correlator by $(1-2e)^2$, giving closed-form expressions for $S(e)$ and $Q(e)$ that produce the simulated violation window to within $0.11$ percentage points and account for its symmetry about $e = 1/2$. Combining the noise model with an explicit multi-pass Cascade reconciliation and the Devetak-Winter bound
gives $r \geq 0.162$ bits per round at the operating point and a positive-rate threshold of $e < 0.89\%$, about six times tighter than the Bell-violation threshold of $5.86\%$: certifying nonlocality is necessary but substantially
weaker than certifying a usable key. The present analysis is asymptotic and assumes collective attacks; extending it to the finite-key regime via the
Entropy Accumulation Theorem \cite{dupuis2019entropy}, measuring
$f_{\text{EC}}$ directly from the disclosed parities, and modelling quantum
frequency conversion for long-distance fiber links are natural next steps.

\begin{credits}
\subsubsection{\ackname}
This research is funded by the Vietnam National Foundation for Science and
Technology Development (NAFOSTED) under Grant 102.01-2025.50 and partially
supported by the VNUHCM - University of Information Technology's Scientific
Research Support Fund.

\subsubsection{\discintname}
The authors have no competing interests to declare that are relevant to the
content of this article.
\end{credits}

\bibliographystyle{splncs04}
\bibliography{ref}

@article{bennett1984quantum,
title = {Quantum cryptography: Public key distribution and coin tossing},
journal = {Theoretical Computer Science},
volume = {560},
pages = {7-11},
year = {2014},
note = {Theoretical Aspects of Quantum Cryptography – celebrating 30 years of BB84},
issn = {0304-3975},
doi = {https://doi.org/10.1016/j.tcs.2014.05.025},
author = {Charles H. Bennett and Gilles Brassard}
}

@article{ekert1991quantum,
  title = {Quantum cryptography based on Bell's theorem},
  author = {Ekert, Artur K.},
  journal = {Phys. Rev. Lett.},
  volume = {67},
  issue = {6},
  pages = {661--663},
  numpages = {0},
  year = {1991},
  month = {Aug},
  publisher = {American Physical Society},
  doi = {10.1103/PhysRevLett.67.661},
}

@article{bennett1992quantum,
  title = {Quantum cryptography without Bell's theorem},
  author = {Bennett, Charles H. and Brassard, Gilles and Mermin, N. David},
  journal = {Phys. Rev. Lett.},
  volume = {68},
  issue = {5},
  pages = {557--559},
  numpages = {0},
  year = {1992},
  month = {Feb},
  publisher = {American Physical Society},
  doi = {10.1103/PhysRevLett.68.557},
}

@article{goh2016measurement,
  title={Measurement-device-independent quantification of entanglement for given Hilbert space dimension},
  author={Goh, Koon Tong and Bancal, Jean-Daniel and Scarani, Valerio},
  journal={New Journal of Physics},
  volume={18},
  number={4},
  pages={045022},
  year={2016},
  publisher={IOP Publishing}
}

@article{brunner2014bell,
  title={Bell nonlocality},
  author={Brunner, Nicolas and Cavalcanti, Daniel and Pironio, Stefano and Scarani, Valerio and Wehner, Stephanie},
  journal={Reviews of modern physics},
  volume={86},
  number={2},
  pages={419--478},
  year={2014},
  publisher={APS}
}

@inproceedings{mayers1998quantum,
  title={Quantum cryptography with imperfect apparatus},
  author={Mayers, Dominic and Yao, Andrew},
  booktitle={Proceedings 39th Annual Symposium on Foundations of Computer Science (Cat. No. 98CB36280)},
  pages={503--509},
  year={1998},
  organization={IEEE}
}

@article{acin2007device,
  title={Device-independent security of quantum cryptography against collective attacks},
  author={Ac{\'\i}n, Antonio and Brunner, Nicolas and Gisin, Nicolas and Massar, Serge and Pironio, Stefano and Scarani, Valerio},
  journal={Physical Review Letters},
  volume={98},
  number={23},
  pages={230501},
  year={2007},
  publisher={APS}
}

@article{pearle1970hidden,
  title = {Hidden-Variable Example Based upon Data Rejection},
  author = {Pearle, Philip M.},
  journal = {Phys. Rev. D},
  volume = {2},
  issue = {8},
  pages = {1418--1425},
  numpages = {0},
  year = {1970},
  month = {Oct},
  publisher = {American Physical Society},
  doi = {10.1103/PhysRevD.2.1418}
}

@article{liu2022toward,
  title={Toward a photonic demonstration of device-independent quantum key distribution},
  author={Liu, Wen-Zhao and Zhang, Yu-Zhe and Zhen, Yi-Zheng and Li, Ming-Han and Liu, Yang and Fan, Jingyun and Xu, Feihu and Zhang, Qiang and Pan, Jian-Wei},
  journal={Physical Review Letters},
  volume={129},
  number={5},
  pages={050502},
  year={2022},
  publisher={APS}
}

@article{gisin2010proposal,
  title = {Proposal for Implementing Device-Independent Quantum Key Distribution Based on a Heralded Qubit Amplifier},
  author = {Gisin, Nicolas and Pironio, Stefano and Sangouard, Nicolas},
  journal = {Phys. Rev. Lett.},
  volume = {105},
  issue = {7},
  pages = {070501},
  numpages = {4},
  year = {2010},
  month = {Aug},
  publisher = {American Physical Society},
  doi = {10.1103/PhysRevLett.105.070501},
}

@article{nadlinger2022experimental,
  title={Experimental quantum key distribution certified by Bell's theorem},
  author={Nadlinger, David P and Drmota, Peter and Nichol, Bethan C and Araneda, Gabriel and Main, Dougal and Srinivas, Raghavendra and Lucas, David M and Ballance, Christopher J and Ivanov, Kirill and Tan, EY-Z and others},
  journal={Nature},
  volume={607},
  number={7920},
  pages={682--686},
  year={2022},
  publisher={Nature Publishing Group UK London}
}

@article{woodhead2021device,
  title={Device-independent quantum key distribution with asymmetric CHSH inequalities},
  author={Woodhead, Erik and Ac{\'\i}n, Antonio and Pironio, Stefano},
  journal={Quantum},
  volume={5},
  pages={443},
  year={2021},
  publisher={Verein zur F{\"o}rderung des Open Access Publizierens in den Quantenwissenschaften}
}

@article{tan2020advantage,
  title={Advantage distillation for device-independent quantum key distribution},
  author={Tan, Ernest Y-Z and Lim, Charles C-W and Renner, Renato},
  journal={Physical Review Letters},
  volume={124},
  number={2},
  pages={020502},
  year={2020},
  publisher={APS}
}

@article{dupuis2019entropy,
  title={Entropy accumulation with improved second-order term},
  author={Dupuis, Fr{\'e}d{\'e}ric and Fawzi, Omar},
  journal={IEEE Transactions on information theory},
  volume={65},
  number={11},
  pages={7596--7612},
  year={2019},
  publisher={IEEE}
}

@misc{githubGitHubDuyndh98DIQKD,
	author = {Nguyen Duong Hoang Duy},
	title = {{G}it{H}ub - duyndh98/{D}{I}{Q}{K}{D}: {D}evice-{I}ndependent {Q}uantum {K}ey {D}istribution - github.com},
	howpublished = {\url{https://github.com/duyndh98/DIQKD.git}},
	year = {2026},
}

@article{navascues2007bounding,
  title={Bounding the set of quantum correlations},
  author={Navascu{\'e}s, Miguel and Pironio, Stefano and Ac{\'\i}n, Antonio},
  journal={Physical Review Letters},
  volume={98},
  number={1},
  pages={010401},
  year={2007},
  publisher={APS}
}

@article{knill2018quantum,
  title={Quantum probability estimation for randomness with quantum side information},
  author={Knill, Emanuel and Zhang, Yanbao and Fu, Honghao},
  journal={arXiv preprint arXiv:1806.04553},
  year={2018}
}

@article{pironio2009device,
  title={Device-Independent Quantum Key Distribution Secure Against Collective Attacks},
  author={Pironio, Stefano and Ac{\'\i}n, Antonio and Massar, Serge and de La Giroday, A. Boyer and Matsukevich, D. N. and Maunz, Peter and Olmschenk, S. and Hayes, D. and Luo, L. and Manning, T. A. and Monroe, C.},
  journal={New Journal of Physics},
  volume={11},
  number={4},
  pages={045021},
  year={2009}
}

@article{vazirani2014fully,
  title={Fully Device-Independent Quantum Key Distribution},
  author={Vazirani, Umesh and Vidick, Thomas},
  journal={Physical Review Letters},
  volume={113},
  number={14},
  pages={140501},
  year={2014}
}

@article{renner2008security,
  title={Security of Quantum Key Distribution},
  author={Renner, Renato},
  journal={International Journal of Quantum Information},
  volume={6},
  number={1},
  pages={1--127},
  year={2008}
}

@article{colbeck2011free,
  title={Free Randomness Can Be Amplified},
  author={Colbeck, Roger and Renner, Renato},
  journal={Nature Physics},
  volume={8},
  number={6},
  pages={450--454},
  year={2012}
}

@article{hensen2015loophole,
  title={Loophole-Free Bell Inequality Violation Using Electron Spins Separated by 1.3 Kilometres},
  author={Hensen, B. and Bernien, H. and Dr{\'e}au, A. E. and Reiserer, A. and Kalb, N. and Blok, M. S. and Ruitenberg, J. and Vermeulen, R. F. L. and Schouten, R. N. and Abell{\'a}n, C. and others},
  journal={Nature},
  volume={526},
  pages={682--686},
  year={2015}
}

@article{giustina2015significant,
  title={Significant-Loophole-Free Test of Bell's Theorem with Entangled Photons},
  author={Giustina, M. and Versteegh, M. A. M. and Wengerowsky, S. and Handsteiner, J. and Hochrainer, A. and Phelan, K. and Steinlechner, F. and Kofler, J. and Larsson, J.-{\AA}. and Abell{\'a}n, C. and others},
  journal={Physical Review Letters},
  volume={115},
  number={25},
  pages={250401},
  year={2015}
}

@article{shalm2015strong,
  title={Strong Loophole-Free Test of Local Realism},
  author={Shalm, Lynden K. and Meyer-Scott, Evan and Christensen, Bradley G. and Bierhorst, Peter and Wayne, Michael A. and Stevens, Martin J. and Gerrits, Thomas and Glancy, Scott and Hamel, Daniel R. and Allman, Michael S. and others},
  journal={Physical Review Letters},
  volume={115},
  number={25},
  pages={250402},
  year={2015}
}

@article{tupkary2023using,
  title = {Using Cascade in quantum key distribution},
  author = {Tupkary, Devashish and L\"utkenhaus, Norbert},
  journal = {Phys. Rev. Appl.},
  volume = {20},
  issue = {6},
  pages = {064040},
  numpages = {13},
  year = {2023},
  month = {Dec},
  publisher = {American Physical Society},
  doi = {10.1103/PhysRevApplied.20.064040},

}

@article{Sekatski2021deviceindependent,
  doi = {10.22331/q-2021-04-26-444},
  title = {Device-independent quantum key distribution from generalized {CHSH} inequalities},
  author = {Sekatski, Pavel and Bancal, Jean-Daniel and Valcarce, Xavier and Tan, Ernest Y.-Z. and Renner, Renato and Sangouard, Nicolas},
  journal = {{Quantum}},
  issn = {2521-327X},
  publisher = {{Verein zur F{\"{o}}rderung des Open Access Publizierens in den Quantenwissenschaften}},
  volume = {5},
  pages = {444},
  month = apr,
  year = {2021}
}

@article{Masini2022simplepractical,
  doi = {10.22331/q-2022-10-20-843},
  title = {Simple and practical {DIQKD} security analysis via {BB}84-type uncertainty relations and {P}auli correlation constraints},
  author = {Masini, Michele and Pironio, Stefano and Woodhead, Erik},
  journal = {{Quantum}},
  issn = {2521-327X},
  publisher = {{Verein zur F{\"{o}}rderung des Open Access Publizierens in den Quantenwissenschaften}},
  volume = {6},
  pages = {843},
  month = oct,
  year = {2022}
}

@article{Tan2021,
  author = {Tan, Ernest Y.-Z. and Schwonnek, René and Goh, Koon Tong and Primaatmaja, Ignatius William and Lim, Charles C.-W.},
  year = {2021},
  title = {Computing secure key rates for quantum cryptography with untrusted devices},
  journal = {npj Quantum Information},
  pages = {158},
  volume = {7},
  number = {1},
  isbn = {2056-6387},
  doi = {10.1038/s41534-021-00494-z},
}

@article{zhang2022device,
  title={A device-independent quantum key distribution system for distant users},
  author={Zhang, Wei and van Leent, Tim and Redeker, Kai and Garthoff, Robert and Schwonnek, Ren{\'e} and Fertig, Florian and Eppelt, Sebastian and Rosenfeld, Wenjamin and Scarani, Valerio and Lim, Charles C-W and others},
  journal={Nature},
  volume={607},
  number={7920},
  pages={687--691},
  year={2022},
  publisher={Nature Publishing Group UK London}
}

@article{clauser1969proposed,
  title = {Proposed Experiment to Test Local Hidden-Variable Theories},
  author = {Clauser, John F. and Horne, Michael A. and Shimony, Abner and Holt, Richard A.},
  journal = {Phys. Rev. Lett.},
  volume = {23},
  issue = {15},
  pages = {880--884},
  numpages = {0},
  year = {1969},
  month = {Oct},
  publisher = {American Physical Society},
  doi = {10.1103/PhysRevLett.23.880},

}

@misc{portmann2014cryptographicsecurityquantumkey,
      title={Cryptographic security of quantum key distribution}, 
      author={Christopher Portmann and Renato Renner},
      year={2014},
      eprint={1409.3525},
      archivePrefix={arXiv},
      primaryClass={quant-ph},
}

@article{tomamichel2011leftover,
  title={Leftover hashing against quantum side information},
  author={Tomamichel, Marco and Schaffner, Christian and Smith, Adam and Renner, Renato},
  journal={IEEE Transactions on Information Theory},
  volume={57},
  number={8},
  pages={5524--5535},
  year={2011},
  publisher={IEEE}
}

@article{Tan2022improveddiqkd,
  doi = {10.22331/q-2022-12-22-880},
  title = {Improved {DIQKD} protocols with finite-size analysis},
  author = {Tan, Ernest Y.-Z. and Sekatski, Pavel and Bancal, Jean-Daniel and Schwonnek, Ren{\'{e}} and Renner, Renato and Sangouard, Nicolas and Lim, Charles C.-W.},
  journal = {{Quantum}},
  issn = {2521-327X},
  publisher = {{Verein zur F{\"{o}}rderung des Open Access Publizierens in den Quantenwissenschaften}},
  volume = {6},
  pages = {880},
  month = dec,
  year = {2022}
}

@article{Primaatmaja2023securityofdevice,
  doi = {10.22331/q-2023-03-02-932},
  title = {Security of device-independent quantum key distribution protocols: a review},
  author = {Primaatmaja, Ignatius W. and Goh, Koon Tong and Tan, Ernest Y.-Z. and Khoo, John T.-F. and Ghorai, Shouvik and Lim, Charles C.-W.},
  journal = {{Quantum}},
  issn = {2521-327X},
  publisher = {{Verein zur F{\"{o}}rderung des Open Access Publizierens in den Quantenwissenschaften}},
  volume = {7},
  pages = {932},
  month = mar,
  year = {2023}
}

@article{Schwonnek2021,
  author = {Schwonnek, René and Goh, Koon Tong and Primaatmaja, Ignatius W. and Tan, Ernest Y.-Z. and Wolf, Ramona and Scarani, Valerio and Lim, Charles C.-W.},
  year = {2021},
  title = {Device-independent quantum key distribution with random key basis},
  journal = {Nature Communications},
  pages = {2880},
  volume = {12},
  number = {1},
  isbn = {2041-1723},
  doi = {10.1038/s41467-021-23147-3},
}

\end{document}